# Log Analysis Techniques using Clustering in Network Forensics

Imam Riadi[1]
[1]Department of Information System, Faculty of
Mathematics and Natural Science,
Ahmad Dahlan University,
Yogyakarta, Indonesia
imam_riadi@uad.ac.id

Jazi Eko Istiyanto[2], Ahmad Ashari[2], Subanar[3]
[2]Department of Computer Science and Electronics,
[3]Department of Mathematics,
[2,3]Faculty of Mathematics and Natural Sciences,
Gadjah Mada University, Yogyakarta, Indonesia
{jazi,ashari}@ugm.ac.id, subanar@yahoo.com

*Abstract* — **Internet crimes are now increasing. In a row with many crimes using information technology, in particular those using Internet, some crimes are often carried out in the form of attacks that occur within a particular agency or institution. To be able to find and identify the types of attacks, requires a long process that requires time, human resources and utilization of information technology to solve these problems. The process of identifying attacks that happened also needs the support of both hardware and software as well. The attack happened in the Internet network can generally be stored in a log file that has a specific data format. Clustering technique is one of methods that can be used to facilitate the identification process. Having grouped the data log file using K-means clustering technique, then the data is grouped into three categories of attack, and will be continued with the forensic process that can later be known to the source and target of attacks that exist in the network. It is concluded that the framework proposed can help the investigator in the trial process.**

## I. INTRODUCTION

Together with the rapidity of internet network development, there are countless individual and business transactions conducted electronically. Communities use the Internet for many purposes including communication, email, transfer and sharing file, search for information as well as online gaming. Internet network offers users to access information that is made up of various organizations. Internet development can be developed to perform digital crimes through communication channels that can not be predicted in advance. However, development of the Internet also provides many sources of digital crime scene. Internet crime is now increasing [1], for example, employees accessing websites that promote pornography or illegal activities that pose a problem for some organizations. Pornography has become a huge business and caused many problems for many organizations. Not only easily available on the Internet but perpetrators also frequently spreading pornography using the advances of Internet technology to attack computer with unsolicited email and pop up ads that are not desirable. Some form of pornography is not only illegal but also bring a big problem for digital investigators. However posting child pornography on the Internet can help lead investigators to the victim. As well as threatening letters, fraud, intellectual property theft is a crime that leaves a digital footprint [2].

Cyber crime, a crime using information technology as instrument or target, have led to the birth of network forensic in response to the rise of the case. Improving the quality of tools and techniques for network forensic analysis is needed to deal with cyber criminals that are more and more sophisticated. Digital forensics, in essence, answer the question: when, what, who, where, how and why related to digital crime [3]. In conducting an investigation into the computer system as an example: when referring to the activity observed to occur, what activities related to what is done, who related to the person in charge, where related to where the evidence is found, how related to activities conducted and why, the activities related to why the crime was committed. Legal regulation of criminal act in the field of information technology is arranged in Law No 11 of 2008 that contains about information and electronic technologies (ITE) contained the provisions of the criminal act elements or the acts that are prohibited in the field of ITE, such as in Article 27, 28, 29, 30, 31, 32, 33, 34, 35 and Article 36. Currently, Indonesian government and House of Representatives are processing on the Information Technology Crime Bill that is included in 247 list of Prolegnas Bill, 2010-2014 [4].

Consequence with many crimes using information technology particularly using the Internet, some crimes are often carried out in the form of attacks that occur within a particular agency or institution. To find and identify the types of attacks, requires a long process that requires time, human resources and utilization of information technology to solve these problems. The process of identifying attacks that happened also needs the support of both hardware and software as well. The attack happened in the Internet network can generally be stored in a log file that has a specific data format. To simplify the process of analyzing the log, the use of scientific methods to help a diverse group of raw data is needed. Clustering technique is one of methods that can be used to help facilitate the identification process.

## II. CURRENT STUDIES ON NETWORK FORENSICS

### A. Forensics in Computer Security

The rapidity of information technology development especially in the field of computer network has brought a positive impact that make human activity becomes easier, faster and cheaper. However, behind all the conveniences it was the development of such infrastructure services have a negative impact emerging in cyberspace, among others: the theft of data on the site, information theft, financial fraud to the Internet, carding, hacking, cracking, phishing, viruses, cybersquating and cyberporn. Some crimes, especially that are using of information technology services spesifically the Internet network can be used to perform some illegal activities that harm others, such as: cyber gambling, cyber terrorism, cyber fraud, cyber porn, cyber smuggling, cyber narcotism, cyber attacks on critical infrastructure, cyber blackmail, cyber threatening, cyber aspersion, phishing.

The number of computer crime cases and computer related crime that is handled by Central Forensic Laboratory of Police Headquarters at around 50 cases, the total number of electronic evidence in about 150 units over a period of time as it can be shown in Table 1. [5].

Table 1. The number of computer crimes and computer related crime cases

| year | number of cases |
|---|---|
| 2006 | 3 cases |
| 2007 | 3 cases |
| 2008 | 7 cases |
| 2009 | 15 cases |
| 2010 (May) | 27 cases |

The forensic process began has been introduced since long time. Several studies related to the forensic process include [5]:

a) Francis Galton (1822-1911); conducted the research on fingerprints
b) Leone Lattes (1887-1954); conducted the research on blood groups (A, B, AB & O)
c) Calvin Goddard (1891-1955); conducted the research on guns and bullets (Ballistic)
d) Albert Osborn (1858-1946); conducted the research on document examination
e) Hans Gross (1847-1915); conducted scientific research on the application of the criminal investigation
f) FBI (1932); conducted the research using Forensic Lab

The forensic process requires a few tools that can help perform forensic processes, Some computer forensic software are shown in Table 2.

Table 2. Forensic Computer Tools

| No | Software | Information |
|---|---|---|
| 1 | E-Detective | http://www.edecision4u.com/ |
| 2 | Burst | http://www.burstmedia.com/release/advertisers/geo_faq.htm |
| 3 | Chkrootkit | http://www.chkrootkit.org |
| 4 | Cryptcat | http://farm9.org/Cryptcat/ |
| 5 | Enterasys Dragon | http://www.enterasys.com/products/advanced-security-apps/index.aspx |
| 6 | MaxMind | http://www.maxmind.com |
| 7 | netcat | http://netcat.sourceforge.net/ |
| 8 | NetDetector | http://www.niksun.com/product.php?id=4 |
| 9 | NetIntercept | http://www.sandstorm.net/products/netintercept |
| 10 | NetVCR | http://www.niksun.com/product.php?id=3 |
| 11 | NIKSUN Function Appliance | http://www.niksun.com/product.php?id=11 |
| 12 | NetOmni | http://www.niksun.com/product.php?id=1 |
| 13 | Network Miner | http://sourceforge.net/projects/networkminer/ |
| 14 | rkhunter | http://rkhunter.sourceforge.net/ |
| 15 | Ngrep | http://ngrep.sourceforge.net/ |
| 16 | nslookup | http://en.wikipedia.org/wiki/Nslookup |
| 17 | Sguil | http://sguil.sourceforge.net/ |
| 18 | Snort | http://www.snort.org/ |
| 19 | ssldump | http://ssldump.sourceforge.net/ |
| 20 | tcpdump | http://www.tcpdump.org |
| 21 | tcpxtract | http://tcpxtract.sourceforge.net/ |
| 22 | tcpflow | http://www.circlemud.org/~jelson/software/tcpflow/ |
| 23 | truewitness | http://www.nature-soft.com/forensic.html |
| 24 | OmniPeek | http://www.wildpackets.com/solutions/network_forensics |
| 25 | Whois | http://www.arin.net/registration/agreements/bulkwhois |
| 26 | Wireshark | http://www.wireshark.org/ |
| 27 | Kismet | http://www.kismetwireless.net/ |
| 28 | Xplico | http://www.xplico.org/ |

CERT defines the forensic as the process of collecting, analyzing, and presenting evidence scientifically in court. Computer forensics is a science to analyze and present data that have been processed electronically and stored in computer media [1]. Digital forensics is the use of scientific methods of preservation, collection, validation, identification, analysis, interpretation, documentation and presentation of digital evidence derived from digital sources or proceeding to facilitate the reconstruction of the crime scene [6].

Indonesia has a state law that can be used to help confirm that crime committed using information technology services may be subject to Article 5 of Law no. 11/2008 on Information and Electronic Transactions (UU ITE) states that electronic information and or electronic documents and or prints with a valid legal evidence can be used as guidelines for processing the crime to the courts, the mechanism of digital evidence uses as adapted to the rules of evidence contained in the investigation.

A few incidents of crimes that often occur in the computer [2]. Digital evidence is defined as the evidentiary value of information stored or transmitted in digital form [7]. A potential source of digital evidence has been growing in the field of mobile equipment [8], Gaming console [9], and digital media devices [10]. Other unique properties of digital evidence is that it can be duplicated. As a result, the evidence must be stored properly at the time of the analysis performed on the copy or copies to ensure that the original evidence was accepted in court [11].

B.  Internet Forensics

American law enforcement agencies began working together in addressing the growing of digital crime in late 1980 and early 1990. Rapid growth of Internet technologies along with increasing volume and complexity of digital crime makes the need for network forensics Internet becomes more important. A state which is not expected to change the future given the number of incidents increased steadily. Figure 1. claimed an increasing number of incidents reported by CERT. [1]

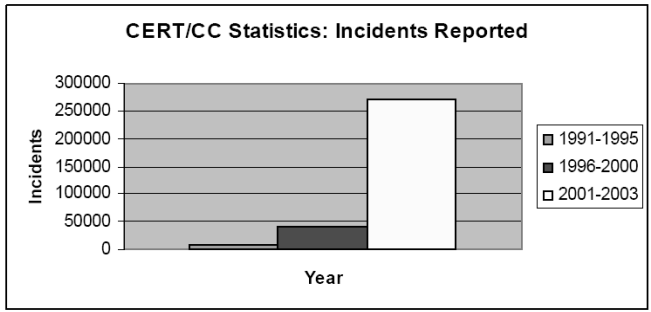

Figure 1. Report the number of incidents by the CERT

C.  Network Forencics

Network forensics is an attempt to prevent attacks on the system and to seek potential evidence after an attack or incident. These attacks include probing, DoS, user to root (U2R) and remote to local.

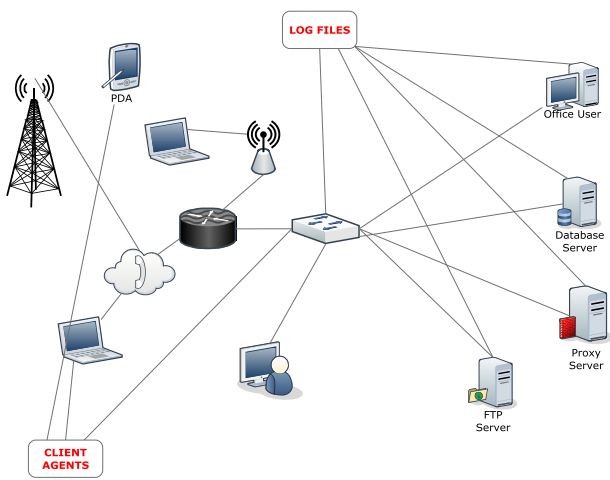

Figure 2. Picture of network forensics process

Figure 2 provides an overview of a network forensics process that occurs within an organization [12]. Network forensics is the process of capturing, recording and analyzing network activity to find digital evidence of an assault or crimes committed against, or run using a computer network so that offenders can be prosecuted according to law [12]. Digital evidence can be identified from a recognizable pattern of attack, deviation from normal behavior or deviations from the network security policy that is applied to the network. Forensic Network has a variety of activities and techniques of analysis as an example: the analysis of existing processes on IDS [13], analysis of network traffic [14] and analysis of the network device itself [15], all of them are considered as the part of network forensics.

Digital evidence can be gathered from various sources depend on the needs and changes in the investigation. Digital evidence can be collected at the server level, proxy level or some other source. For example the server level digital evidence can be gathered from web server logs that store browsing behavior activities that are frequented. The log describes the user who access the website and what are they do. Several sources including the contents of network devices and traffic through both wired and wireless networks. For example, digital evidence can be gathered from the data extracted by the packet sniffer like: tcpdump to monitor traffic entering the network [16].

### III. THEORETICAL BACKGROUND

A.  Network Abnormal Detection in Computer Security

Anomaly detection refers to the problem of finding patterns in data that are inconsistent with expected behavior. Patterns that do not fit often called as an abnormal condition that often occurs within a network. The detection of abnormal tissue can be found in several applications such as credit card fraud detection, insurance or health care, intruder detection for network security, fault detection is critical to the system as well as observations on the military to find enemy activity. Anomaly detection can translate the data in significant so way that it can present information that is useful in various application domains. For example, the presence of abnormal patterns that occur in network traffic that can be interpreted that the hacker sends sensitive data for unauthorized purposes [17].

B.  The concept of Network Abnormal Detection

Anomaly patterns in the data that do not fit well with the notion of normal behavior. Figure 3 depicts anomalies in a simple 2-dimensional data that have been defined, which has two normal regions, N1 and N2, because the most frequent observation in a two-way areas [17]. Examples of points O1 and O2, and O3 point in the region, are the anomalies.

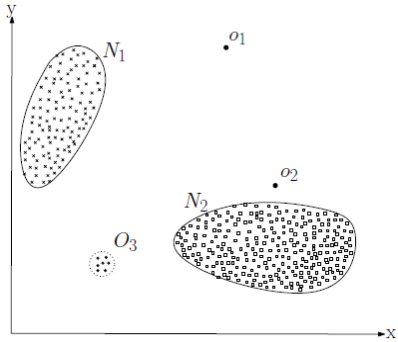

Figure 3. a simple example of an anomaly in the data 2-dimensional.

Anomaly may be caused by many things, for example malicious activities, like credit card fraud, terrorist activities or making hang the system, but all reason have common characteristics that it is interesting to be analyzed. Above caused most of the abnormal is not easy to solve. Most of the abnormal detection techniques can solve these problems. Detection of abnormal has become a major topic in research, [18] among others provides a broad survey of the abnormal detection techniques are developed using machine learning and statistical domains. Review techniques for detection of abnormal numerical data by [19]. Review of detection techniques using neural networks and statistical approaches by [20] and [21].

**C. Clustering**

Clustering is a process to make the grouping so that all members of each partition has a certain matrix equation based on [22]. A cluster is a set of objects that were merged into one based on equality or proximity. Clustering as a very important technique that can perform translational intuitive measure of equality into a quantitative measure. Here is an example of the clustering process as shown in Figure 4 [22].

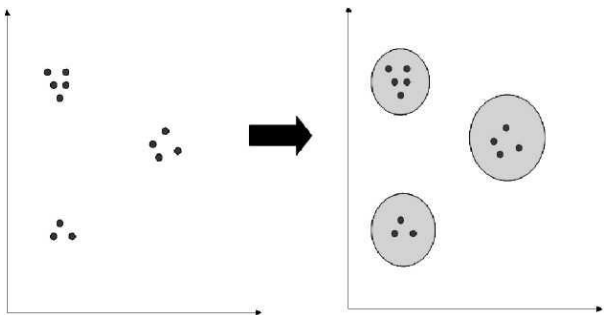

Figure 4. Clustering based on proximity

Figure 4. is an example of the process of clustering the data using proximity as a parameter. The data that are near will be clustered each other as a member of the cluster. Clustering characteristics can be grouped into 4 types as described below :

a) Partitioning clustering
 Partitioning clustering is also called exclusive clustering, where each data must belong to a particular cluster. Characteristics of this type also allow for any data that includes a specific cluster in a process step, the next step moving to another cluster.
 Example: K-Means, residual analysis.
b) Hierarchical clustering
 In the hierarchical clustering, every data must belong to a particular cluster, and the data that belongs to a particular cluster at a stage of the process can not move to another cluster at a later stage.
 Example: Single Linkage, Centroid Linkage, Complete Linkage, Average Linkage.
c) Overlapping clustering
 In overlapping clustering, each data allows belong to multiple clusters. The data has a value of membership (membership) in a cluster.
 Example: Fuzzy C-means clustering, Gaussian Mixture.
d) Hybrid
 Hybrid characteristics is the cluster characteristics that combines the characteristics of the clustering characteristics of the partitioning, overlapping, and hierarchical

Grouping method is basically divided into two, namely the method of grouping hierarchy (Hirarchical Clustering Method) and the method of Non Hierarchy (Non Hirarchical Clustering Method). Hierarchical clustering method is used when no information on the number of groups to be selected. While the non-hierarchical clustering method aims to classify objects into k groups (k <n), where the value of k has been determined previously. One of the Non Hierarchical clustering procedure is to use K-Means method. This method is a method of grouping which aims to group objects so that the distance of each object to the center of the group within a group is the minimum [22].

**D. K-Means Clustering**

K-means is included in the partitioning clustering that also called exclusive clustering separates the data into k separate parts and each of the data should belong to a particular cluster and allows for any data that includes a specific cluster in a process step, the move to the next stage cluster other [22]. K-means is algorithm that is very famous because of its ease and ability to perform the grouping of the data and outliers of data very quickly. In the K-means any data should be included into a specific cluster, but allows for any data that includes a specific cluster in a process step, the next step moving to another cluster. Figure 5 shows illustration of the process steps clustering using K-means algorithm [22] as follows :

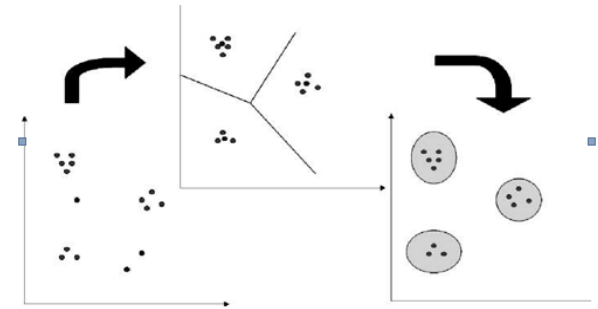

Figure 5. Illustration of the process steps clustering using K-means algorithm.

K-Means algorithm on clustering can be done by following these steps [22]:
a) Determine the number of clusters k to be formed.
b) Generate k centroids (cluster center) beginning at random.
c) Calculate the distance of each data to each centroid.
d) Each data choose the nearest centroid.
e) Determine new centroid position by calculating the average value of the data that choose the same centroid.
f) Return to step 3 if the new centroid position is not same with the old centroid.

Here are the advantages of K-means algorithm in the clustering process [22]:
a) K-means is very fast in the clustering process.
b) K-means is very sensitive to the random generation of initial centroid.
c) Allows a cluster has no members
d) The results of clustering with K-means is not unique (always changing), sometimes good, sometimes bad
e) K-means is very difficult to reach the global optimum

Moreover, K-means algorithm has a drawback that the clustering results are very dependent on the initialization initial centroids that are randomly generated, and therefore allows for any particular cluster of data that includes a process step, the next stage move to another cluster. In the net stage Figure 6 illustrates the weakness of K-means algorithm showed that in the previous stages there are three clusters with a cluster which do not have any member and on the next stage there is cluster formation that is just consist of two cluster and all of them have members [22], of course this is caused by the centroid that is operated at random.

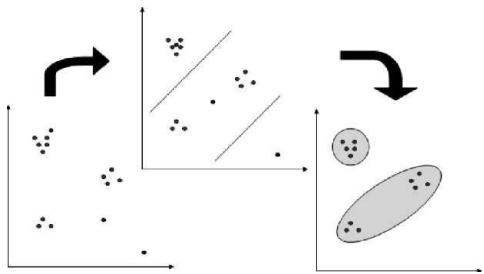

Figure 6. Illustration of K-means algorithm weakness.

## IV. CASE STUDY

Topology that used in this research aims to facilitate the investigation process is shown in Figure 7.

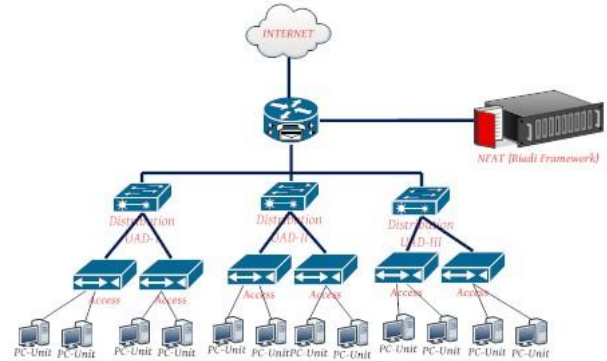

Figure 7. The design of topology research

Framework Module NFAT (Network Forensic Analysis Tool) is developed using open source software that can run on any operating system platform, among others (Linux, Unix, FreeBSD, OpenBSD), this application was developed with shell scripting, combined with PHP and supported using the MySQL DBMS.
Experiments and testing framework NFAT module is done at the Center for Computer Laboratory Ahmad Dahlan University, Yogyakarta, to obtain the appropriate data for the data traffic flowing in a computer network is large enough.

This research will be developed using a framework that is shown in Figure 8

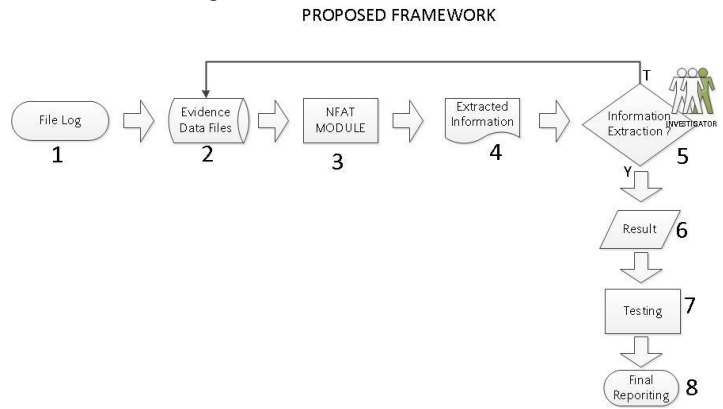

Figure 8. Model Framework to be developed

In Figure 8. First-stage of forensic process starting from the collection of evidence collected in connection with the initial written by the investigators as evidence profiles and the input to the database of evidence, evidence management system sought by finding the appropriate case-related data and time. In the analysis phase, the input data generated by the log file system, then the database will be stored in evidence. When the investigator and the investigator needs information, the information extracted from Module NFAT (Network Forensic Analysis Tools). At the investigation stage, the extracted information is considered as part of the investigation.

Although it is very fast final decision depends on the investigator. Investigator will determine whether the evidence has been produced to meet or not, if the evidence has not been met, it will be back again to extract data from evidence database. Otherwise if the evidence meets the test process will be done to verify that the data is original and suitable with the criteria of evidence that required by investigators. In the final stage of reporting, digital evidence will be presented in a particular format so that it can help the investigator in the trial process.

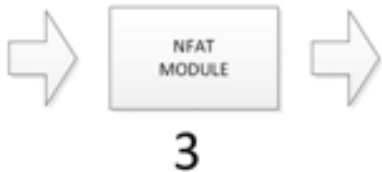

Figure 9. Framework Module NFAT

NFAT module as shown in Figure 9 works using K-means clustering algorithm which can perform the detection of attacks based on grouping the data into three groups of attacks, namely [22]:
a) dangerous attack,
b) rather dangerous attack,
c) not dangerous attack.

Based on the data stored in the database log file system, then the clustering process will be done in stages as follows [22]:
a) Specified value of k as the number of clusters to be formed.
b) Generate k centroids (cluster center) beginning at random.
c) Calculate the distance of each data to each centroid.
d) Each data choose the nearest centroid.
e) Determine new centroid position by calculating the average value of the data that choose the same centroid.
f) Return to step c if the new centroid position is not same with the old centroid.

The results of the data cluster for an attack is highly dependent on the generation of its centroid because it is done at random, this resulted in the detection of an attack on the data is always changing. Once the data clustering process is carried out the attack, then each cluster results do cluster labeling is included in the hazard, rather dangerous or not dangerous. Then from the cluster that has been labeled, checked against is done against the data which are entered into the next group of malicious attacks on the note in the report. The process of clustering using K-means algorithm is shown in Figure 10 [22].

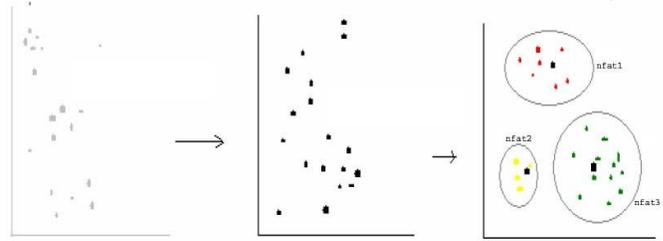

Figure 10. The process of clustering the data with the K-means attack

From the data mentioned above cluster that are formed is the best cluster obtained from the cluster that has the smallest variance. Of the above forms clusters, each cluster for the data had been formed but has not been labeled, the labeling is done by calculate for the matrix multiplication of the final centroid of each cluster is multiplied by its transpose matrix so we get a scalar value of each cluster, as shown in Table 3 [22].

Table 3. Cluster grouping type of attack

| No | Cluster | ID |
|----|---------|----|
| 1 | nfat1 | 1,3,6,7,10,16 |
| 2 | nfat2 | 9,11,12,13 |
| 3 | nfat3 | 2,4,8,14,15,17 |

From the result of transpose multiplication each centroid of three cluster above for example the results obtained with the sequence results from the largest to the small cluster nfat1, nfat2 and cluster nfat3 cluster, The cluster having the highest transpose multiplication result would be labeled as the dangerous cluster. So that the matrix multiplication of the cluster was obtained by labeling the cluster nfat1 is a malicious attack, an attack cluster is somewhat harmful nfat2 and nfat3 is not dangerous cluster attack [22].

In addition it has done in module development framework NFAT (Network Forensic Analysis Tool) to facilitate the forensic process is carried out in accordance with the Internet network research plan that has been made. Here are some of the infrastructure supporting the development of NFAT module framework to facilitate the process of forensic analysis of Internet network. The following log data extracted from the database used to identify the attack as shown in Figure 11.

Figure 11. The data used to perform classification of attacks

The module output data of NFAT is a clustering process, where the results of this cluster can be calculated error values to be compared with the target data that is the target of the cluster. The target data used for comparison are shown in Table 4 [22].

Tabel 4. List of criteria attack

| Protocol | Criteria | Port | TCPFlag |
|---|---|---|---|
| TCP | dangerous attack | 80,8080,443 20,21 22,23 | 16,32 |
| TCP | Rather dangerous attack | 161,143,162, 110,993 | The combination of binary digits 20-24 |
| TCP | not dangerous attack | In addition to the above mentioned | The combination of binary digits 20-27 |
| UDP | dangerous attack | 53 | - |
| UDP | Rather dangerous attack | 137,161, | - |
| UDP | not dangerous attack | In addition to the above mentioned | - |

Having grouped the data log file using K-means clustering technique, then the data is grouped into 3 categories of attack, and then will resume the forensic process that can later be known to the source and target of the attack on the network, this type of attack which occurs on TCP (Transmission Control Protocol) is shown in Figure 12.

Figure 12. The data that perform the types of attacks occurred on the TCP protocol.

The type of attack that occurred in the UDP (User Datagram protocol) can be shown in figure 13.

Figure 13. The data that perform the types of attacks occurred on the UDP protocol.

## V. CONCLUSIONS

The first stage of the forensic process starting from collection of evidence which is collected in connection with the initial case that is written by the investigators as evidence profiles and entries to the evidence database, evidence management system is sought by finding the appropriate case related data and time. In the analysis phase, the input data generated by the log file system, then the data will be stored in evidence database. When the investigators need information, the information extracted from Module NFAT (Network Forensic Analysis Tools). At the investigation stage, the extracted information is considered as the part of the investigation. Although that process is very fast, the final decision depends on the investigator. Then the investigator will determine whether the evidence that is produced has been met or not, if the evidence has not been met, it will back again to the extract data from evidence database, otherwise if the evidence has been met, the test process will be done to verify that the data is original and appropriate with the criteria of evidence that is needed by investigator. In the final stage of reporting, digital evidence will be presented in a particular format so that it can help the investigator in the trial process.

### ACKNOWLEDGMENT

The authors would like to thank Ahmad Dahlan University (http://www.uad.ac.id) that provides funding for the research, and the Department of Computer Science and Electronics Gadjah Mada University (http://mkom.ugm.ac.id) that provides technical support for the research.

### REFERENCES

[1] CERT, CERT/CC Statistics 1988-2005, CERT-Research-Annual-Report. (http: //www .cert. org/stats), 2008

[2] Kruse II, W.G. and Heiser, J.G. Computer forensics: incident response essentials. Addison-Wesley, 2002

[3] Beebe, N.L. and Clark, J.G. A hierarchical, objectives-based framework for the digital investigations process. Proceedings of the fourth Digital Forensic Research Workshop. 2004

[4] Syamsuddin A, Tindak Pidana Khusus, Sinar Grafika, Jakarta, 2011
[5] Alamsyah R, Digital Forensic, Security Day 2010, Inixindo, Yogyakarta, 2010.
[6] SWGDE, Best Practices for Computer Forensics, Scientific Working Group on Digital Evidence, 2007.
[7] Pollitt, M.M. Report on digital evidence. Proceedings of the Thirteenth International Forensic Science Symposium, 2001
[8] Mellars, B. Forensic examination of mobile phones. Digital Investigation, vol. 1, no. 4, pp. 266-272, 2004
[9] Vaughan, C. Xbox security issues and forensic recovery methodology (utilising Linux). Digital Investigation, vol. 1, no. 3, pp. 165-172. 2004
[10] Marsico, C.V. and Rogers, M.K. iPod forensics. International Journal of Digital Evidence, vol. 4, no. 2. 2005
[11] Meyers, M. and Rogers, M. Computer forensics: the need for standardization and certification. International Journal of Digital Evidence, vol. 3, no. 2. 2004
[12] Mukkamala, S. and Sung, A.H. Identifying significant features for network forensic analysis using artificial techniques. International Journal of Digital Evidence, vol. 1, no. 4. 2003
[13] Sommer, P. Intrusion detection systems as evidence. Computer Networks, vol. 31, no. 23-24, pp. 2477-2487. 1999
[14] Casey, E. Handbook of computer crime investigation: forensic tools and technology. Academic Press. 2004
[15] Petersen, J.P. Forensic examination of log files. MSc thesis, Informatics and Mathematical Modelling, Technical University of Denmark, Denmark. 2005
[16] Jacobson, TCPDump-dump traffic on a network. Retrieved February, 2006
[17] Chandola.V, Banerjee.A, Kumar.V, Anomaly Detection : A Survey, A modifed version of this technical report will appear in ACM Computing Surveys, 2009
[18] Hodge, V. and Austin, J. A survey of outlier detection methodologies. Artificial Intelligence Review 22, 2, 85-126. 2004
[19] Agyemang M, Barker K, Alhaj R, A comprehensive survey of numeric and symbolic outlier mining techniques. Intelligent Data Analysis 10, 6, 521 538, 2006
[20] Markou, M. and Singh, S. Novelty detection: a review-part 1: statistical approaches.Signal Processing 83, 12, 2481 2497. 2003a
[21] Markou, M. and Singh, S. Novelty detection: a review-part 2: neural network based approaches. Signal Processing 83, 12, 2499 2521. 2003b
[22] Fauziah L, Computer Network Attack Detection Based on Snort IDS with K-means Clustering Algorithm, ITS Library, 2009

**AUTHORS PROFILE**

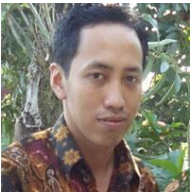

Imam Riadi is a lecturer of the Bachelor Information System Program, Matematics and Natural Science Faculty of Ahmad Dahlan University Yogyakarta, Indonesia. He was graduated as S.Pd. in Yogyakarta State University, Indonesia. He got his M.Kom. in Gadjah Mada University, Indonesia. He is currently taking his Doctoral Program at the Computer Science and Electronics Department of Gadjah Mada University Yogyakarta, Indonesia.

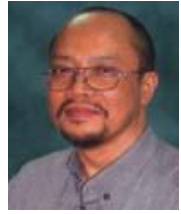

Jazi Eko Istiyanto is a Professor and the Head of Computer Science and Electronics Department, Universitas Gadjah Mada Yogyakarta, Indonesia. He holds a B.Sc in Physics from Gadjah Mada University, Indonesia. He got his Postgraduate Diploma (Computer Programming and Microprocessor), M.Sc (Computer Science) and PhD (Electronic System Engineering) from University of Essex, UK.

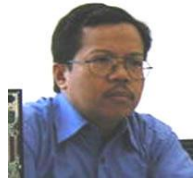

Ahmad Ashari is an Associate Professor at Computer Science and Electronics Department of Gadjah Mada University Yogyakarta, Indonesia. He was graduated as Bachelor of Physics in Gadjah Mada University, Indonesia. He received his M.Kom. in Computer Science in University of Indonesia, and received his Dr. techn. in Informatics at Vienna University of Technology, Austria.

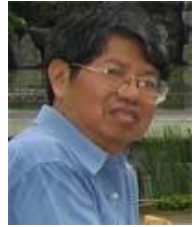

Subanar is a Professor at the Department of Mathematics, Gadjah Mada University in Yogyakarta, Indonesia. He was graduated as Bachelor of Mathematics from Gadjah Mada University and Ph.D (Statistics) at Wisconsin University, USA.